\documentclass[9pt,conference]{IEEEtran}

\usepackage{bm}
\usepackage{multirow}
\usepackage{multicol}
\usepackage{makecell}
\usepackage{listings}

\usepackage{dcase2025,amsmath,graphicx,url,times,booktabs, tabularx}

\title{MIMII-Agent: Leveraging LLMs with Function Calling for \\Relative Evaluation of Anomalous Sound Detection}

\name{Harsh Purohit, Tomoya Nishida, Kota Dohi, Takashi Endo, and Yohei Kawaguchi}
\address{Hitachi Ltd., R\&D Group, Tokyo, Japan}

\begin{document}

\maketitle

\begin{abstract}
This paper proposes a method for generating machine-type-specific anomalies to evaluate the relative performance of unsupervised anomalous sound detection (UASD) systems across different machine types, even in the absence of real anomaly sound data. 
Conventional keyword-based data augmentation methods often produce unrealistic sounds due to their reliance on manually defined labels, limiting scalability as machine types and anomaly patterns diversify. 
Advanced audio generative models, such as MIMII-Gen, show promise but typically depend on anomalous training data, making them less effective when diverse anomalous examples are unavailable.  
To address these limitations, we propose a novel synthesis approach leveraging large language models (LLMs) to interpret textual descriptions of faults and automatically select audio transformation functions, converting normal machine sounds into diverse and plausible anomalous sounds. 
We validate this approach by evaluating a UASD system trained only on normal sounds from five machine types, using both real and synthetic anomaly data. 
Experimental results reveal consistent trends in relative detection difficulty across machine types between synthetic and real anomalies. 
This finding supports our hypothesis and highlights the effectiveness of the proposed LLM-based synthesis approach for relative evaluation of UASD systems.
\end{abstract}


\section{Introduction}
Detecting anomalies in machine sounds is a critical aspect of predictive maintenance, aiming to prevent equipment failures and reduce downtime in industrial operations~\cite{koizumi2020description}. 
Machine sounds provide valuable insights into equipment health, where anomalies may indicate issues such as mechanical wear, misalignment, or impending component failure.  
Traditional anomaly detection methods~\cite{foggia2015audio,coelho2022deep, suefusa2020anomalous} rely heavily on data, but collecting extensive datasets of real anomalous sounds is challenging due to the rarity and unpredictability of faults. 
Deliberately inducing faults is often impractical or hazardous, leading to a fundamental scarcity of real anomalous data.

This scarcity and lack of diversity in available real anomaly data significantly hinder not only training but also the evaluation of anomalous sound detection (ASD) systems across diverse fault conditions. 
Rigorous assessment of system capabilities requires suitable test data, which is often unavailable or fails to adequately represent real-world conditions. 
While conventional data augmentation~\cite{inoue2020detection, wilkinghoff2023fraunhofer, chen2023effective} and text-to-audio (TTA)-based anomaly generation~\cite{zahedi2023regularized, zhang2024first, purohit2024mimii} attempt to address this, they face limitations: synthesized sounds may lack realism, or advanced generative models may require anomalous samples for training, further complicating evaluation when real-world anomaly data are sparse or absent. 
Consequently, reliably benchmarking ASD performance against diverse, realistic fault conditions across different machine types remains a significant challenge. 

To address this issue, we first introduce the concept of \textbf{``relative evaluation''}, which complements conventional absolute evaluation. 
Relative evaluation assesses a system’s comparative strengths and weaknesses by verifying \textbf{detection performance rankings across machine types}, where the system performs well for some machine types but poorly for others.
This approach is essential in heterogeneous industrial environments, where maintenance engineers must know where a system is trustworthy and where it is not, enabling optimized sensor allocation, inspection scheduling, and risk management. 
Traditional absolute metrics, such as the Area Under the ROC Curve (AUC), can fluctuate with the severity of anomaly samples and become unreliable when real anomaly data are scarce. 
Specifically, severe anomalies in the test set lead to easy detection and high AUC scores, while mild anomalies make detection harder, resulting in lower AUC scores.
In contrast, relative evaluation focuses on the comparative difficulty of detection tasks across machine types, providing consistent insights even when anomaly severity varies.

We further propose a novel method for synthesizing anomalous sounds to enable relative evaluation of ASD systems across machine types.  
The method leverages LLMs’ implicit world knowledge---i.e., ``common sense''---to interpret textual descriptions of faults and automatically apply appropriate acoustic transformations to normal machine sounds, thereby generating plausible acoustic characteristics for the described faults.
This approach generates diverse, controllable synthetic anomalies without requiring prior anomalous data or manual intervention, facilitating relative evaluation of ASD systems' strengths and weaknesses across machine types.

\section{Related research}

\subsection{Unsupervised anomalous sound detection}

Unsupervised Anomalous Sound Detection (UASD) is critical for ASD, especially when anomalous data is unavailable. 
The DCASE challenges~\cite{koizumi2020description, kawaguchi2021description, dohi2022description, dohi2023description, nishida2024description} have advanced this field by providing datasets like ToyADMOS series~\cite{koizumi2019toyadmos, harada2021toyadmos, harada2023toyadmos, niizumi2024toyadmos} and MIMII series~\cite{purohit2019mimii, tanabe2021mimii, dohi2022mimii}, enabling benchmarking of techniques like autoencoder-based methods~\cite{koizumi2018unsupervised}, Gaussian-Mixture-Model-based methods~\cite{mnasri2021anomalous}, embedding-similarity-based approaches~\cite{wang2021unsupervised}, etc. 
Despite these advancements, evaluating detection systems across diverse fault conditions remains difficult due to the lack of representative anomalous test data, highlighting the need for synthetic datasets that simulate realistic anomalies.

\subsection{Anomalous sound generation}

\begin{table*}[t]
\centering
\caption{Conventional Approaches for TTA-based Anomalous Sound Generation}
\label{tab:related}
\begin{tabular}{lcccc}
\toprule
\textbf{Characteristic} & Zahedi et al.~\cite{zahedi2023regularized} & Zhang et al.~\cite{zhang2024first} & MIMII-Gen~\cite{purohit2024mimii} & \textbf{This work} \\
\midrule
\textbf{Machine-type-aware realistic synthesis} & No & 
Yes (uses metadata) & Yes (uses metadata) & \textbf{Yes (uses metadata)} \\[0.6em]
\textbf{Trainable with normal data alone}  & Yes & No & No & \textbf{Yes} \\[0.6em]
\textbf{Purpose} & Training & Training & Evaluation (absolute) & \textbf{Relative evaluation} \\
\bottomrule
\end{tabular}
\end{table*}

Synthetic data generation addresses the shortage of real anomalous sounds for training and evaluation.
Most conventional methods for anomalous sound generation focus on data augmentation to train UASD systems using anomalous examples.

One common approach is basic data augmentation, which includes pitch-shifting and time-stretching~\cite{inoue2020detection}, Mixup-based augmentation~\cite{wilkinghoff2023fraunhofer}, and statistics-exchange-based augmentation~\cite{chen2023effective}.
However, these methods often produce unrealistic sounds that are unsuitable for robust evaluation.

Another approach is TTA-based anomalous sound generation (See Table \ref{tab:related}).
Zahedi et al.'s method~\cite{zahedi2023regularized} generates anomalous sounds by randomly selecting prompts created by ChatGPT and feeding them into AudioLDM~\cite{liu2023audioldm}. However, this method cannot achieve realistic, machine-type-specific synthesis because it does not consider machine type when selecting prompts.
Zhang et al.'s method~\cite{zhang2024first} converts metadata into captions, which are then input into AudioLDM. This approach can potentially achieve realistic synthesis by leveraging metadata such as machine type, but it requires anomalous samples for training.
MIMII-Gen~\cite{purohit2024mimii} has been proposed specifically for evaluating anomaly detection systems.
Similar to Zhang et al.~\cite{zhang2024first}, MIMII-Gen converts metadata into captions and uses these captions as input for a diffusion model to generate anomalous sounds. While it can potentially achieve machine-type-aware realistic synthesis, it also requires anomalous samples for training, which limits its applicability when such data are sparse.

\subsection{TTA models}
This subsection highlights representative TTA models, which serve as the foundation for the above TTA-based anomaly-generation approaches~\cite{zahedi2023regularized, zhang2024first, purohit2024mimii}.
TTA models leverage the contextual understanding of LLMs to generate speech, music, or environmental sounds directly from textual prompts.
For high-fidelity generation, latent diffusion models have demonstrated exceptional effectiveness. AudioLDM~\cite{liu2023audioldm} pioneered CLAP~\cite{elizalde2023clap}-conditioned latent diffusion, enabling zero-shot audio generation.
Building on similar approaches, recent works have achieved improvements in multi-domain synthesis quality and enhanced temporal coherence~\cite{huang2023make, huang2023make2, liu2024audioldm2}.
To address inference latency, some models compress diffusion into fewer steps~\cite{liu2024audiolcm, bai2024consistencytta, saito2025soundctm}.
TANGO 2~\cite{majumder2024tango2} employs preference optimization to align audio outputs with human-perceived prompt consistency, using reinforcement-style feedback to train more reliable generators.
However, none of the above TTA models are trained on industrial machine recordings, while Zhang et al.~\cite{zhang2024first} and MIMII-Gen~\cite{purohit2024mimii} explicitly train their generators on machine-type-labeled industrial sound datasets, enabling type-specific fault synthesis.

\subsection{Research gap and contribution}
As mentioned earlier, traditional absolute metrics commonly used in anomaly detection benchmarks~\cite{koizumi2020description, kawaguchi2021description, dohi2022description, dohi2023description, nishida2024description} fluctuate with the severity of anomaly samples, making them unreliable when real anomaly data are scarce. 
Severe anomalies in the test set lead to easy detection and high AUC scores, while mild anomalies make detection harder, resulting in lower AUC scores.

To address this limitation, our first contribution is the introduction of relative evaluation, which verifies detection performance rankings across machine types. 
Unlike absolute evaluation, which is sensitive to anomaly severity, relative evaluation identifies where the system performs better or worse.

Our second contribution is a scalable method for generating diverse and realistic anomalous sounds using LLMs to enable relative evaluation. 
Instead of directly generating synthetic anomalies or manually adding them, our approach leverages LLMs to interpret machine types and their corresponding anomalies, transforming normal machine audio into anomalous audio. 
As shown in Table \ref{tab:related}, unlike the conventional TTA-based anomalous sound generation methods~\cite{zahedi2023regularized, zhang2024first, purohit2024mimii}, our approach enables realistic synthesis tailored to specific machine types and operates with training only on normal data. 
This ensures reliable evaluation across different machines, even when real anomaly data are limited.

\section{Proposed method}

\begin{figure*}[t]
\centering 
\includegraphics[height=2.6 in , width=6.5 in]{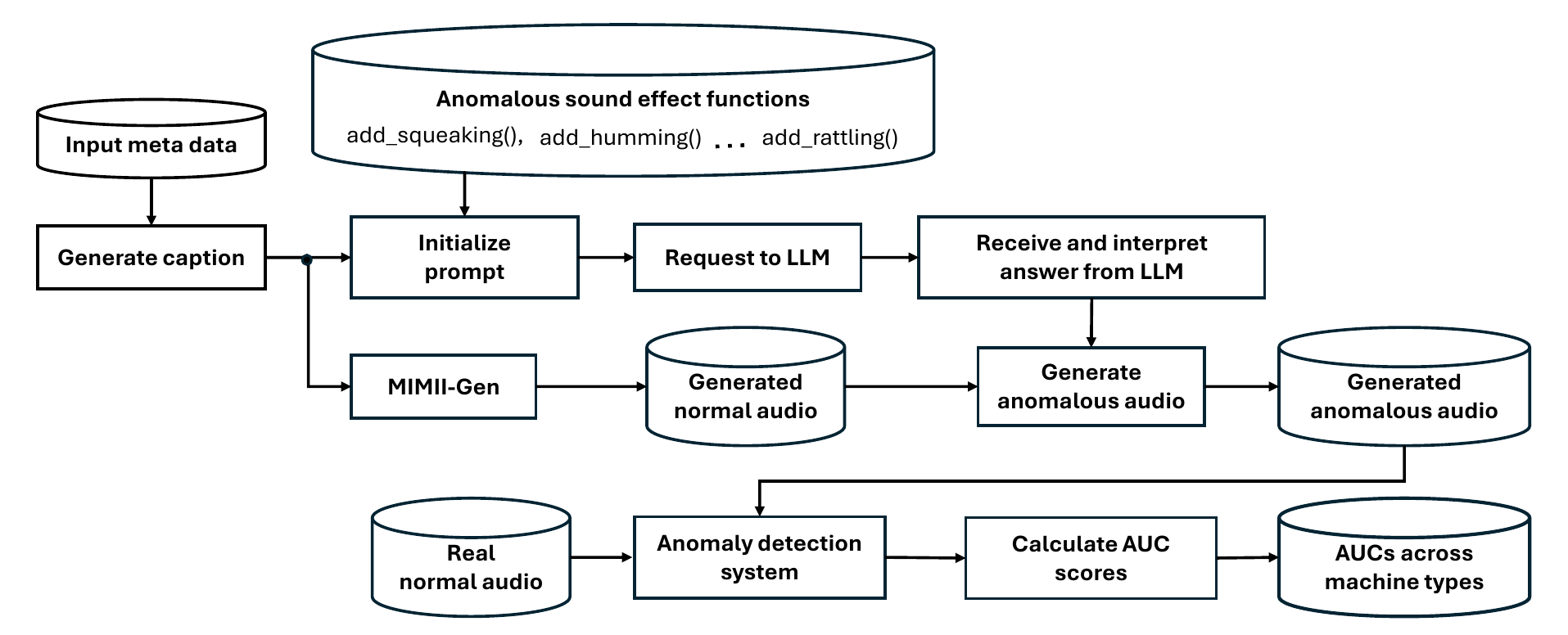} 
\caption{Workflow of Our Approach} 
\label{fig:workflow} 
\end{figure*}
Our proposed method introduces a novel approach to generating anomalous machine sounds by leveraging LLMs to intelligently select and apply appropriate sound effects to normal machine audio based on descriptive captions. The comprehensive workflow illustrated in Fig.~\ref{fig:workflow} consists of two major parts: (a) synthetic anomalous sound generation and (b) relative evaluation of anomaly detection systems across different machine types.

\subsection{Synthetic anomaly sound generation}
\subsubsection{Workflow}
The synthetic anomaly generation process follows a systematic workflow as illustrated in the block diagram:
\begin{itemize}
\item \textbf{Input metadata:} The process begins with metadata that provides contextual information about machine types, operating and environmental conditions.
\item \textbf{Generate caption:} Based on the input metadata, the system generates descriptive caption using Flan-T5~\cite{chung2024scaling} that characterize the machine's operational state.
\item \textbf{Generate normal audio:}
Using the MIMII-Gen latent diffusion model \cite{purohit2024mimii}, we generate high-fidelity normal machine audio that serve as the foundation for anomaly introduction.
\item \textbf{Initialize prompt:} This step consists of carefully crafted system prompt, user prompt containing generated caption and anomalous sound effect functions described in section \ref{functions}.
\item \textbf{Request to LLM:} In this step, the initialized prompt is sent to a Large Language Model via an API call. The language model analyzes the captions and autonomously selects the most appropriate sound effect to simulate potential anomaly relevant to the operating condition present in the caption. 
\item \textbf{Receive and interpret answer from LLM:} The system parses the LLM's response, extracts the selected function name, and maps it to the corresponding audio processing function from a predefined library of anomalous sound effects.
\item \textbf{Anomalous audio generation:} The selected function is applied to the normal audio obtained from MIMII-Gen, transforming it into anomalous audio with context-appropriate fault characteristics. The generated anomalous sounds are stored along with the applied anomaly effects. 
\end{itemize}

\subsubsection{Anomalous sound effect functions}
\label{functions}

We implement a comprehensive library of sound effect functions that simulate various machine fault conditions:
\begin{itemize}
    \item \textbf{Squeaking or Squealing}: High-pitched sounds indicating faulty bearings or friction between components.
    \item \textbf{Rattling or Knocking}: Noises suggestive of loose parts or misalignment.
    \item \textbf{Grinding or Scraping}: Sounds indicative of severe mechanical wear or damage.
    \item \textbf{Humming or Buzzing}: Low-frequency sounds resulting from electrical issues or resonance.
    \item \textbf{Whistling or Hissing}: Sounds associated with air leaks or high-pressure flow.
    \item \textbf{Clicking or Tapping}: Noises caused by relay switches or intermittent contacts.
    \item \textbf{Pulsing or Chattering}: Effects indicating fluctuating power supply or control issues.
    \item \textbf{Pop or Bang}: Sudden sounds simulating abrupt failures or explosive events.
    \item \textbf{Changes in Tonal Quality or Frequency}: Alterations representing shifts in machine operation.
    \item \textbf{Broadband Noise Increases}: Overall noise level increases to simulate general degradation.
\end{itemize}
Each function uses specific digital signal processing techniques to modify normal audio waveforms, creating realistic fault signatures.

\textbf{Example:} The sound effect functions are implemented using Python libraries such as NumPy, Librosa, and SoundFile, manipulating the audio waveforms to introduce the selected anomalies.
\begin{lstlisting}[language=Python, basicstyle=\footnotesize\ttfamily]
def add_squeaking(audio, sr, 
    duration=2.0, freq=4000, intensity=0.3):
    # Adds a high-pitched squeaking sound 
    # to the audio
    # Function implementation
    return audio_with_squeaking
\end{lstlisting}

\subsection{Anomaly detection system evaluation}

\subsubsection{Unsupervised anomaly detection system}
The UASD system is trained exclusively on real normal audio and calculates anomaly scores, such as reconstruction errors in autoencoders, to identify anomalies. 
This unsupervised approach aligns with real-world scenarios where anomalies are rare and underrepresented in training data.  
The UASD system can utilize methods such as autoencoder-based~\cite{koizumi2018unsupervised}, Gaussian Mixture Model-based~\cite{mnasri2021anomalous}, or embedding-similarity-based approaches~\cite{wang2021unsupervised}. 
The system processes all clips from real normal sound and synthetic anomaly datasets and computes anomaly scores for each clip.

\subsubsection{AUC calculation and relative evaluation}
The AUC score for each machine type is calculated as:
\begin{equation}
{\rm AUC}_{m} = \frac{1}{N^{-}_{m}N^{+}_{m}} \sum_{i=1}^{N^{-}_{m}} \sum_{j=1}^{N^{+}_{m}}
\mathcal{H} (\mathcal{A}(x_{j}^{+}) - \mathcal{A}(x_{i}^{-})),
\end{equation}
where $m$ represents the machine type index,  
$\mathcal{H}(x)$ returns 1 if $x > 0$ and 0 otherwise,  
$\{x^{-}_{i}\}_{i=1}^{N^{-}_{m}}$ are normal test clips,  
and $\{x_{j}^{+}\}_{j=1}^{N^{+}_{m}}$ are anomalous test clips for machine type $m$.  
$N^{-}_{m}$ and $N^{+}_{m}$ indicate the number of normal and anomalous test clips, respectively.

By comparing AUC scores across machine types, users can identify the system’s relative strengths and weaknesses based on detection performance rankings.

\section{Experimentation}
The experiments aim to validate two key objectives.  
The first objective is to confirm the correlation between synthetic and real anomaly detection performance rankings across different machine types, demonstrating that synthetic anomaly generation using our proposed approach enables users to identify the system’s relative strengths and weaknesses.  
The second objective is to validate the effectiveness of LLM-based approaches in generating contextually appropriate synthetic anomalies compared to manual and random methods, through an ablation study.  
This section provides details on the dataset, anomaly detection system, results, and an ablation study.

\subsection{Dataset preparation}
Table \ref{tab:dataset_summary} summarizes the datasets used in this study.  
We prepared three distinct datasets to support our evaluation: normal sounds, synthetic anomalous sounds, and real anomalous sounds. 
Each dataset was tailored to the five machine types under study.

Table \ref{tab:dataset_summary} summarizes the datasets used in this study.  
We prepared three distinct datasets tailored to five machine types: bearings, gearboxes, fans, valves, and slide rails. 
\begin{itemize}
    \item \textbf{Normal Sounds:} We collected 900 normal sound recordings per machine type (bearings, gearboxes, fans, valves, and slide rails), each 10 seconds long, sampled at 16 kHz. These were sourced from industrial environments and public datasets like MIMII-DG \cite{dohi2022mimii} ensuring a comprehensive representation of typical operating conditions.
    \item \textbf{Synthetic Anomalous Sounds:} For each machine type, we generated 50 synthetic anomalies by applying sound effects to normal sounds. The language model (GPT-4) selected effects (e.g., squeaking, rattling) based on captions describing operating conditions (e.g., "Bearing operating at 24 krpm"). 
    \item \textbf{Real Anomalous Sounds:} We acquired 50 real anomalous recordings per machine type. These anomalies represent actual faults, such as mechanical wear or misalignment, and vary in severity.
\end{itemize}
\begin{table}[t]
\centering
\caption{Dataset Summary}
\label{tab:dataset_summary}
\begin{tabular}{lccc}
\toprule
\textbf{Category} & \textbf{Samples per} & \textbf{Duration} & \textbf{Sample Rate} \\
& \textbf{Machine} & & \\
\midrule
Normal Sounds & 900 & 10 s & 16 kHz \\
Synthetic Anomalies & 50 & 10 s & 16 kHz \\
Real Anomalies & 50 & 10 s & 16 kHz \\
\bottomrule
\end{tabular}
\end{table}

\subsection{Anomaly detection system design}

We employed an unsupervised anomaly detection system based on an autoencoder, trained exclusively on normal sounds to detect deviations indicative of anomalies.

\textbf{Autoencoder Architecture:}
\begin{itemize}
    \item \textbf{Input Layer:} Log-mel spectrograms with 128 mel-bins, extracted from 64-ms frame windows with 50\% hop size.
    \item \textbf{Encoder:} Three layers (128, 64, 32 filters, kernel size 3), each followed by ReLU activation and max-pooling (2x2).
    \item \textbf{Decoder:} Mirrored layers with upsampling, reconstructing the input spectrogram.
\end{itemize}

\textbf{Training Details:} The autoencoder was trained for 100 epochs with a batch size of 32, using the Adam optimizer (learning rate 0.001) and mean squared error (MSE) loss. A single model was trained across all machine types to generalize normal patterns, reflecting real-world scenarios with diverse equipment.

\subsection{Results}

\begin{table}[t]
\centering
\caption{Detection Performance on Synthetic vs. Real Anomalies}
\label{tab:results}
\begin{tabular}{lcccccc}
\toprule
\multirow{3}{*}{\makecell[l]{\textbf{Machine}\\\textbf{Type}}} 
& \multicolumn{3}{c}{\textbf{Synthetic}} 
& \multicolumn{3}{c}{\textbf{Real}} \\
\cmidrule(lr){2-4} \cmidrule(lr){5-7}
& \multicolumn{2}{c}{\textbf{AUC}} & \multirow{2}{*}{\textbf{Rank}} 
& \multicolumn{2}{c}{\textbf{AUC}} & \multirow{2}{*}{\textbf{Rank}} \\
& \textbf{MSE} & \textbf{MAHALA} & & \textbf{MSE} & \textbf{MAHALA} & \\
\midrule
Bearing    & 0.85 & 0.82 & 3 & 0.57 & 0.61 & 3 \\
Gearbox    & 0.88 & 0.86 & 2 & 0.62 & 0.67 & 2 \\
Fan        & 0.92 & 0.95 & 1 & 0.90 & 0.93 & 1 \\
Slide rail & 0.80 & 0.79 & 4 & 0.55 & 0.57 & 4 \\
Valve      & 0.78 & 0.72 & 5 & 0.53 & 0.52 & 5 \\
\bottomrule
\end{tabular}
\end{table}

Table \ref{tab:results} summarizes the detection performance for synthetic and real anomalies using AUC scores, derived from two distinct metrics: Mean Squared Error (MSE) and Mahalanobis distance (MAHALA). Employing both scoring methods provides a more robust validation of the relative evaluation on synthetic data.
Synthetic anomalies consistently achieved higher AUC scores  compared to real anomalies, indicating they are easier to detect.  
Notably, the ranking of AUC scores, and thus the relative anomaly detection difficulty, is consistent across machine types for both synthetic and real data. 
These results demonstrate that synthetic anomaly generation using our proposed approach effectively enables users to evaluate the system’s relative strengths and weaknesses.

Differences in AUC scores reflect the unique operational behaviors and fault characteristics of each machine type. 
For example, lower AUC scores for valves and slide rails suggest that anomalies in these machines may be subtler or involve features that are harder for the anomaly detection system to capture.

\subsection{Ablation study}

To validate the reliability of the LLM-based approach, we compared three configurations: 
(1) our approach with GPT-4o-based anomaly function calling (with values reproduced from Table \ref{tab:results}),
(2) a keyword-based manual mapping of anomalies created through human labeling based on common knowledge (e.g., adding a "squeaking" anomaly if the caption includes "bearing"), and (3) random selection of possible anomalies without contextual understanding. 

\begin{table}[t]
\centering
\caption{Ablation Study: Common Knowledge Impact on MSE-AUC}
\label{tab:ablation_lm}
\begin{tabular}{p{40pt}cccccc}
\toprule
\multirow{3}{*}{\makecell[l]{\textbf{Machine}\\\textbf{Type}}} & \multicolumn{2}{c}{\textbf{GPT-4o}} & \multicolumn{2}{c}{\textbf{Manual-mapping}} & \multicolumn{2}{c}{\textbf{Random}} \\
 & & & \multicolumn{2}{c}{\textbf{(w/ Knowledge)}} & \multicolumn{2}{c}{\textbf{(w/o Knowledge)}} \\
\cmidrule(lr){2-3}\cmidrule(lr){4-5}\cmidrule(lr){6-7}
 & AUC & rank & AUC & rank & AUC & rank \\
\midrule
Bearing & 0.85 & 3 & 0.70 & 3 & 0.81 & 5 \\
Gearbox & 0.88 & 2 & 0.72 & 2 & 0.85 & 3 \\
Fan & 0.92 & 1 & 0.78 & 1 & 0.82 & 4 \\
Slide rail & 0.80 & 4 & 0.67 & 4 & 0.86 & 2 \\
Valve & 0.78 & 5 & 0.65 & 5 & 0.89 & 1 \\
\bottomrule
\end{tabular}
\end{table}
Table \ref{tab:ablation_lm} presents the results of the ablation study.  
The AUC-score rankings across machine types produced using GPT-4o and the keyword-based manual mapping approach formed by human labels were both closely aligned with those of real anomalies.  
In contrast, random anomaly selection showed no correlation with the AUC-score rankings of real anomalies, highlighting the importance of contextual understanding in anomaly generation. 
These results confirm that the ability of LLMs to interpret machine-specific characteristics and fault descriptions, leveraging common knowledge embedded within them, enables the creation of realistic and relevant anomalies. 
Furthermore, the observation that sound effects impact machines differently emphasizes the need to tailor the selection process to each machine type.  
The consistent performance of LLM-based anomaly generation demonstrates its potential as a scalable and efficient alternative to human labeling, capable of supporting diverse machine types and anomaly scenarios. 
Improving prompt design and incorporating domain-specific constraints could further enhance the realism of generated anomalies, thereby increasing the approach's utility for UASD evaluation.

\section{Conclusion}

This paper addressed the challenge of evaluating UASD systems in the absence of sufficient and diverse real anomaly data.  
To tackle this, we proposed two key contributions:
(1) the introduction of relative evaluation, which verifies detection performance rankings across machine types. 
Unlike absolute evaluation, which is sensitive to anomaly severity, relative evaluation identifies where the system performs better or worse. 
(2) a novel synthesis approach using LLMs with function-calling capabilities. 
Our method leverages the world knowledge of LLMs to interpret textual descriptions of machine conditions and automatically apply audio transformations, generating diverse and plausible synthetic anomalies for evaluation, without requiring prior anomalous examples.  

Experiments showed that AUC-score rankings are consistent across machine types as well as different anomaly detection systems for both synthetic and real data.
Also, rankings produced using GPT-4o and keyword-based manual mapping closely aligned with those of real anomalies, while random anomaly selection showed no correlation, highlighting the importance of contextual understanding in anomaly generation. 
These findings validate the ability of LLMs to generate realistic and relevant anomalies by interpreting machine-specific characteristics and fault descriptions. 
Our approach offers a reliable and scalable tool for benchmarking UASD performance and understanding relative detection difficulty across machine types, especially in scenarios lacking sufficient real anomaly data.


\bibliographystyle{IEEEtran}
\bibliography{refs}







\end{document}